\documentclass[aps,prb,twocolumn,groupedaddress,showpacs,showkeys]{revtex4}
\usepackage{graphics,graphicx}
\usepackage{color}
\usepackage{amssymb}
\usepackage{bookmath}
\usepackage{bm}

\begin{document}
\title{Vortex state in $d$-wave superconductors with strong paramagnetism:
transport and specific heat anisotropy}
\author{A. B. Vorontsov}
\affiliation{Department of Physics,
             Montana State University, Bozeman, Montana, 59717, USA}
\author{I. Vekhter}
\affiliation{Department of Physics and Astronomy,
             Louisiana State University, Baton Rouge, Louisiana, 70803, USA}
\date{\today}

\begin{abstract}
We analyse the combined effect of orbital and Pauli depairing on
the superconducting state, and apply the results to the heavy
fermion CeCoIn$_5$. We find that: a) standard extrapolation
based on the slope of $H_{c2}(T)$ in
the vicinity of the transition temperature to $T=0$ does not
always give accurate values of the orbital upper critical field;
b) critical value of the Maki parameter, $\alpha$, that determines
onset of the first order transition depends on the Fermi surface
shape and the symmetry of the gap, and is $\alpha^*\approx 3$ for
CeCoIn$_5$; c) the anisotropy of the thermodynamic and transport
coefficients in the low-temperature, low-field part of the phase
diagram is essentially insensitive to the Zeeman field and can be
used to determine the nodal directions in Pauli-limited
superconductors. The latter result confirms the finding of the
$d_{x^2-y^2}$ order parameter CeCoIn$_5$.
\end{abstract}
\maketitle

\section{Introduction}

Magnetic field is one of the most widely utilized and powerful
probes of the unconventional superconducting state. It couples to
momentum and spin of the quasiparticles, via the orbital and
Zeeman mechanisms respectively, and therefore can be used to probe
both the momentum dependence of the order parameter and the spin
structure of the Cooper pairs.

In singlet superconductors both mechanisms are detrimental to
superconductivity. The orbital coupling of the Cooper pair motion
to the vector potential of the field leads, in type-II
superconductors, to the appearance of a mixed state with partial
field penetration. The Cooper pair supercurrents and the order
parameter are spatially modulated and form an array of Abrikosov
vortices. The upper critical field at which the second order
transition into the normal state occurs is given $H_{c2}^{orb}
\sim \Phi_0 / 2\pi \xi^2$, where $\Phi_0$ is the flux quantum, and
$\xi$ is the superconducting coherence length~\cite{Tinkham}. For
unconventional superconductors, below that field the number of the
quasiparticles outside of the superconducting condensate, and,
consequently, entropy-sensitive properties such as the specific
heat or thermal conductivity, depend on the relative orientation
of the field with respect to zeroes (nodes) or deep minima in the
energy gap
\cite{IVekhter:1999R,IVekhter:2001,HWon:2001,PThalmeier:2002,PMiranovic:2003,NNakai:2004,PMiranovic:2005,AVorontsov:2006,ABVorontsov:2007dos,ABVorontsov:2007kap,GRBoyd:2009}.
The corresponding measurements have been extensively used to
determine the symmetry of the superconducting state
\cite{TPark:2003,TParkReview:2004,HAoki:2004,FYu:1995,HAubin:1997,TWatanabe:UPd2Al3,KIzawa:YNiBCkappa,KIzawa:BEDT,KIzawa:CeCoIn5,YMatsuda:2006,TSakakibara:2007,KYano:2008,KAhn:2009}.
with the heavy fermion CeCoIn$_5$ and CeIrIn$_5$ two examples
where a reasonably detailed comparison of theory and experiment
was carried
out~\cite{KIzawa:CeCoIn5,HAoki:2004,AVorontsov:2006,ABVorontsov:2007dos,ABVorontsov:2007kap}.
Crucially, the anisotropic signal changes sign depending on the
values of $T$ and $H$ in the superconducting phase diagram, so that
either minima or maxima correspond to the direction of the field
along the
nodes~\cite{AVorontsov:2006,ABVorontsov:2007dos,ABVorontsov:2007kap}.
Consequently, a detailed theoretical analysis is required for the
interpretation of the experimental data. So far, such an analysis
has only been carried out for purely orbital coupling.

At the same time measurements on CeCoIn$_5$ clearly show that the
second suppression mechanism, depairing due to Pauli spin
polarization, is important in this compound
\cite{TTayama:2002,ABianchi:2002,ABianchi:Sci08}. From the theory
perspective, magnetic field aligns the spins of the unpaired
electrons, while the singlet Cooper pairs cannot take advantage of
the lower energy offered by a spin-polarized state. As a result,
in the absence of the orbital effects, the normal state is
energetically favorable above the Pauli critical field $H_P(T=0) =
\Delta_s/\sqrt{2}\mu_B$ (for isotropic superconductors, see below),
where $\Delta_s$ is the superconducting gap, and
$\mu_B=\hbar e /2 m_e c$ is the Bohr magneton~\cite{cha62,clo62}.
The normal state is reached via a second order transition at $T >
T^*_P \sim 0.56 \, T_c$ and a first order transition at low
temperatures $0 < T < T^*_P$.  In the absence of strong anisotropy
of the spin-orbit coupling, the number of unpaired electrons at
$H<H_P$ does not depend on the field orientation and therefore
whenever this mechanism is dominant it is not immediately
obvious how efficient the field is for determination of the gap
symmetry, and when exactly the inversion of the anisotropy
pattern takes place.

Under most circumstances the orbital suppression mechanism
dominates, $H_{c2}^{orb}\ll H_P$. Pauli limiting is important in
thin films and other quasi-two-dimensional materials, when the
orbital coupling is
inefficient~\cite{PMTedrow:1970,WWu:1995,PWAdams:1998}. In
strongly layered materials even at relatively low fields the
Zeeman pairbreaking effect may complicate the extraction of the
nodal contribution to the specific
heat~\cite{IVekhter:2001,MIchioka:2007}. Even in relatively 3D
systems, if the characteristic electron velocity (Fermi velocity)
is low and hence the coherence length is short, as it is in many
heavy fermion and other correlated superconductors, the orbital
upper critical field is high, and can be comparable to the Pauli
limiting field. The question of what the vortex state anisotropies
can tell us about the superconducting gap symmetry in this
situation has not been explored before and is the subject of this
paper.

We consider a model relevant to CeCoIn$_5$,
\cite{ABianchi:2002,TTayama:2002,cap04}. The situation when the
orbital and Pauli pairbreaking mechanisms are comparable was
considered early on by Gruenberg and Gunther~\cite{gru66} and by
Maki~\cite{mak66} for an $s$-wave superconductor with spherical
Fermi surface (FS). CeCoIn$_5$ has $d$-wave gap symmetry with the
main $f$-electron containing Fermi surface sheet open along the
$c$-axis~\cite{DHall:2001,RSettai:2001,YHaga:2001,NHarrison:2004}.
Consequently, as our first task, we compute the upper critical
field as a function of temperature for different orientations. We
extend previous calculations~\cite{ada03,ike04,won04a} to the
Fermi surface in the shape of a corrugated cylinder that gives the
correct normal state resistivity anisotropy, and find the the
temperature $T^*$, below which the normal to superconducting
transition becomes first-order. One of our main findings is that
the commonly used criterion for determining the dominant Pauli
limiting regime via the Maki parameter~\cite{mak66}, $\alpha =
\sqrt{2} H^{orb}_{c2}/H_P>1$ is quantitatively incorrect and
depends on the symmetry of the gap and on the shape of the Fermi
surface.

Second, we analyse the anisotropy of the specific heat and the
thermal conductivity under rotated magnetic field across the
$T$-$H$ phase diagram including the effects of Pauli limiting.
Since the precise details of the behavior of the anisotropy are
used to infer the gap symmetry from experiment, this extension is
critical for justifying $d_{x^2-y^2}$ symmetry in CeCoIn$_5$. We
show that even moderately high Pauli effect has little influence
on the the low temperature and low field part of the phase diagram
whereas its consequences near the upper critical field are
considerable.

\section{Quasiclassical formulation}

\subsection{Basic equations for superconductors under Zeeman and
orbital field.}

We use the quasiclassical formalism~\cite{eil68,lar68} at real
frequencies, $\vare$, which allows to carry out calculations for
arbitrary temperature and field and to self-consistently include
effects of the field and impurities on the order parameter. The
quasiclassical transport equation for the matrix Green's function
in particle-hole and spin space, $\whg$, has the form~\cite{ale85}
    \bea [ (\vare + {e\over c}
\vv_f(\hat{\vp}) \vA(\vR) )\, \widehat{\tau}_3 - \mu \vB \cdot
\hat{\vS} - \whDelta(\vR, \hat{\vp}) - \whs_{imp}(\vR; \vare),
 \nonumber \\
\whg(\vR, \hat{\vp}; \vare)]
+ i\vv_f(\hat{\vp}) \cdot \gradR \; \whg(\vR, \hat{\vp}; \vare) = 0 \,. \hspace*{1cm}
\label{eq:eilZ}
    \eea
Here $[E_1, E_2]$ denotes a commutator, we carried out the
standard~\cite{ser83} separation of the center of mass coordinate,
$\vR$, and the momentum of the relative motion, $\vp$, so that the
Fermi
velocity depends on the position on the Fermi surface, 
$\vv_f(\hat{\vp})$. The orbital coupling is via the the electron
charge and the vector potential, $\vA(\vR)$, and the Zeeman term,
$\mu \vB \cdot \hat{\vS}$, is proportional to the electron's
magnetic moment $\mu = (\mathfrak{g}/2) \mu_B$ with
$\mathfrak{g}$-factor as a material-specific parameter, and
the spin matrix,
    \be \hat{\vS} =\left(
\begin{array}{cc} \vsigma & 0 \\ 0 & \vsigma^* \end{array} \right)
\,
    \ee
with $\vsigma$ the usual vector of Pauli matrices. The (retarded)
Green's function in particle-hole and spin space,
    \be \whg^R =
\left(
\begin{array}{cc}
g+\vg\vsigma & (f+\vf\vsigma)i\sigma_y \\
i\sigma_y (f'+\vf' \vsigma) & -g+\vg\vsigma^*
\end{array} \right) \,,
    \ee
satisfies normlization $\whg^2 = -\pi^2 \widehat{1}$.
Eq.~(\ref{eq:eilZ}) is complemented by two other equations. One is
the used to determine the impurity self-energy, which we treat in
the self-consistent $t$-matrix approximation
\cite{PJHirschfeld:1988}, $\whs_{imp}(\vR; \vare) = n_{imp}
\widehat{t}(\vR; \vare)$, with
\begin{equation}
  \widehat{t}(\vR; \vare) = u \widehat{1} +
  u N_0 \langle \whg (\vR, \hat{\vp}; \vare)\rangle_{\hvp} \widehat{t}
  \,,
\end{equation}
where $N_0$ is the density of states at the Fermi level. In
writing this equation we assumed non-magnetic isotropic scattering
with the individual impurity potential strength $u$, so that the
impurity self energy does not have any momentum
dependence~\cite{PJHirschfeld:1988}. The second equation is the
self-consistency condition on the order parameter which relates it
to the off-diagonal, in particle-hole space, component of the
Green's function. Before we write it explicitly, however, it is
convenient to rewrite the function $\whg$ in a different
representation.


Recall that we are interested in the regime $H\gg H_{c1}$, when
the vortices form an Abrikosov lattice, and are considering
strongly type-II superconductors. In this case the internal field
$\vB$ is essentially uniform and equal to the applied field, $\vH$.
The important point is that for a field which orientation is
the same at all spatial points we can choose the spin quantization
axis along the field $\vH=H \hat{\vz}$,
and introduce for all vector quantities the notation
$\vx\cdot\vsigma=\bar{x} \sigma_z$.  Then the $4\times4$ Green's
function, Eq.~(\ref{eq:eilZ}) consists of two blocks,
    \be \whg^R =
\left(
\begin{array}{cccc}
g+\bar{g} & 0 & 0 & f+\bar{f} \\
0 & g-\bar{g} & -f+\bar{f} & 0 \\
0 & f'-\bar{f}' & -g+\bar{g} & 0 \\
-f'-\bar{f}' & 0 & 0 & -g-\bar{g}
\end{array} \right) \,,
\ee corresponding to the spin-up and spin-down components. The
impurity self-energy $\whs_{imp}$ assumes the same block-form.

The block structure allows us to rewrite the equations for the
components of the Green's function in a simple form. We explicitly
introduce the equations for the two spin components, $s =
\{\uparrow(+1),\downarrow(-1)\}$, via $g_\sm{s} = g +s \bar{g}$,
$f_\sm{s} = f + s \bar{f}$. These functions now satisfy
independent quasiclassical equations and normalization conditions,
$g_\sm{s}^2 - f_\sm{s} f'_\sm{s} = -\pi^2$, with the Zeeman energy
shift ${\vare} \to {\vare} \mp \mu B$ for up and down spin
respectively, see also Ref.~\onlinecite{kle04}.
Now, for example, equation for the off-diagonal part of the
Green's function takes the form
 \bea
\left[-2i(\tilde{\vare}_\sm{s}-s\mu B) + \vv_f(\hat{\vp}) \left(
\gradR - i \frac{2e}{\hbar c} \vA(\vR) \right) \right] \, 
f_\sm{s}(\vR, \hat{\vp}; \vare)
\nonumber \\
 = 2 \widetilde{\Delta}_\sm{s}(\vR; \vare) \, i g_\sm{s}(\vR, \hat{\vp}; \vare)
 \,.
 \label{eq:fspin}
\hspace{1cm}
    \eea
Here we introduced the shorthand notations
$\tilde{\vare}_\sm{s}=\vare-\Sigma_\sm{s}$ and
$\widetilde{\Delta}_\sm{s}=\Delta+\Delta_{imp,\sm{s}}$ explicitly
including the diagonal ($\Sigma_s$) and off-diagonal,
$\Delta_{imp,\sm{s}}$ components of $\whs_{imp}$. The spins mix
only through the self-consistency equation for the singlet order
parameter,
    \bea \Delta (\vR, \vk; \vare) &=&\int\limits_{-\infty}^{+\infty} {d\vare
\over 4\pi i } \tanh {\vare\over 2T}
\\
\nonumber &\times& \Big<  V(\vk,\vp)\,
(f_{\uparrow}(\vR,\hvp;\vare)+f_{\downarrow}(\vR,\hvp; \vare)
\Big>_{\hvp} \,,
    \eea
where $V(\vk,\vp)$ is the pairing potential. Eq.~(\ref{eq:fspin})
has the same form as the quasiclassical equation in the absence of
Zeeman term, and therefore we can utilize the existing techniques
to solve for each of the two spin components independently,
enforcing the self-consistency for the order parameter at the
final step.

\subsection{Model and method of solution}

We follow the approach we developed
earlier~\cite{IVekhter:1999,AVorontsov:2006,ABVorontsov:2007dos,ABVorontsov:2007kap}
and solve the equations using a modified Brandt-Pesch-Tewordt
(BPT) approximation~\cite{BPT:1967,WPesch:1975}. We assume the
existence of the Abrikosov vortex lattice and model the spatial
dependence of the order parameter by
$\Delta (\vR, \vp)= \Delta(\hvp) \braket{\vR}{0}
= \Delta \cY(\hvp) \braket{\vR}{0}$,
where $\cY(\hvp)$ is the normalized
($\langle \cY(\hvp)^2\rangle_{\hvp}=1$) basis function for the
irreducible representation corresponding to the chosen gap
symmetry, and normalized by a proper choice of $C_{k_y}$
coefficients the spatial vortex lattice profile,
\begin{equation}
  \braket{\vR}{0}= \sum_{k_y} C_{k_y} {e^{ik_y\sqrt{S_f} y}
    \over \sqrt[4]{S_f \Lambda^2}} \widetilde\Phi_0\left( x,k_y
    \right)\,.
    \label{eq:DeltaPhi}
\end{equation}
The magnetic length $\Lambda^2=hc/2eB$, and
\begin{equation}
  \widetilde\Phi_0(x,k_y)=\Phi_0\left( {x-\Lambda^2
    \sqrt{S_f} k_y\over \Lambda \sqrt{S_f}} \right) \,,
    \label{Phi-0}
\end{equation}
is the ground state oscillator function, and  $x$ and $y$ are in
the direction normal to the field.~\cite{ABVorontsov:2007dos}
Here we approximated the vortex lattice by only the
superposition of the lowest oscillator wave functions,
$\ket{0}$. The admixture of the higher oscillator states is
small~\cite{ILukyanchuk:1987,HWon:1996} and does not substantially
affect the conclusions regarding the properties in the vortex
state~\cite{HWon:1996,AVorontsov:2006,ABVorontsov:2007dos,ABVorontsov:2007kap}.
Hence we consider only the lowest Landau level, as reflected in
Eq.~(\ref{eq:DeltaPhi}), but with properly rescaled Fermi velocity
component perpendicular to the field. For a Fermi surface
rotationally invariant around the $c$ axis that we consider
below~\cite{ABVorontsov:2007dos}
    \be S_f = \left[\cos^2 \theta_H +
    {v_{0||}^2\over v_{0\perp}^2} \sin^2 \theta_H\right]^{1/2}
    \,,
    \ee
$\theta_H$ is the angle between the field direction and the
$c$-axis,  $v_{0\parallel}^2 = 2 \langle \cY^2(\hvp)
v^2_\parallel(p_z) \rangle_\sm{FS}$, and $v_{0\perp}^2 = 2 \langle
\cY^2(\hvp) v^2_{\perp i}(p_z) \rangle_\sm{FS}$, where
$v_\parallel$ is the $c$-axis component of the Fermi velocity,
while $v_{\perp i}$ with $i=a,b$ is the Fermi velocity component
in the $a$-$b$ plane. For the field in the basal plane
$\theta_H=\pi/2$, and therefore $S_f=v_{0||}/ v_{0\perp}$.

The BPT approximation consists of replacing the diagonal part of
the Green's function with its spatial average, and is justified
over a wide range of fields~\cite{EHBrandt:1995,TDahm:2002}. The
equations for the off-diagonal components of the Green's function
are solved by introducing the ladder operators as in
Ref.~\onlinecite{ABVorontsov:2007dos,ABVorontsov:2007kap}, and, in
conjunction with the normalization condition,
for the lowest Landau level give
\bea
 && f_s= \pi \frac{1}{\sqrt{1+P}} \frac{2\sqrt{\pi}\Lambda}{|\tilde{v}_f^\perp|}
  \, W \left[ \frac{2(\tilde{\vare}-s\mu B)\Lambda}{|\tilde{v}_f^\perp|} \right] \,
  \tilde\Delta_s \,,
    \nonumber 
\\
 && g_s= \pi \frac{-i}{\sqrt{1+P}} \,,
    \label{eq:gs}
\\
 && P =  -i\sqrt{\pi}\left(\frac{2\Lambda}{|\tilde{v}_f^\perp|}\right)^2
    \, W^\prime \left[ \frac{2(\tilde{\vare}-s\mu B)\Lambda}{|\tilde{v}_f^\perp|} \right] \,
    \widetilde\Delta_s \ul{\widetilde\Delta}_s \,,
    \nonumber
\eea
with $ \widetilde\Delta_s(\hvp,\vare) = \Delta(\hvp) + \Delta_{imp,s}(\vare)$,
$W(z) = \exp(-z^2) \mbox{erfc}(-iz)$, and
\begin{equation}
  |\tilde{v}_f^\perp|=\left[\frac{v_{f,x}(\hat{\vp})^2}{{S_f}}+
  v_{f,y}^2(\hat{\vp})
  {S_f}\right]^{1/2}\,.
\end{equation}
This closed form solution is used to enforce the self-consistency
on the impurity self-energy and the gap value. The approach gives
the order parameter and the Green's function that can be used to
determine the physical properties below.

\section{Fermi surface, upper critical field, and Maki parameter}

We consider a model Fermi surface that approximates the main sheet
of CeCoIn$_5$ as seen by the magnetic oscillations
~\cite{RSettai:2001,YHaga:2001,HShishido:2002,NHarrison:2004}.
It has the shape of an open cylinder rotationally symmetric in the
$a$-$b$ plane (component of the momentum labeled $p_r$) and
modulated along the $c$-axis ~\cite{AVorontsov:2006}, and is given
by $p^2_f = p_r^2-r^2 p_f^2 \cos(2 a p_c/r^2 p_f)$ with $r=a=0.5$.
This choice of parameters gives a moderate anisotropy between
transport coefficients in the $c$-direction and $a$-$b$ plane, close
to that of CeCoIn$_5$ in the normal state~\cite{AMalinowski:2005}.
With this definition, the typical quasiparticle velocity in the
basal plane is $v_f = p_f/m$, and that along the $c$-axis,
$v_{f,c} \sim a v_f = 0.5 v_f$.

We choose the $d_{x^2-y^2}$ symmetry of the order parameter and
take a model separable pairing interaction
$V(\vk,\vp)=V_0\cY(\hvk)\cY(\hvp)\equiv
V_0\cY(\phi)\cY(\phi^\prime)$, where $\phi$ labels the azimuthal
angle around the Fermi surface, and $\cY(\phi)=\sqrt{2}\cos
2\phi$. With this choice $\Delta(\vR,\hvp) = \sqrt{2} \Delta(\vR)
\cos2\phi$. The dimensionless coupling constant $N_0V_0$
determines the transition temperature of the pure sample,
$T_{c0}$, which is suppressed by impurities to $T_c$; we use
this latter $T_c$ as a unit of energy. Similarly, the
natural unit for the magnetic field is the characteristic orbital
scale, $B_0 = \Phi_0/2\pi\xi_0^2$, where $\Phi_0=hc/2e$ is the
flux quantum, and $\xi_0 = \hbar v_f/2\pi T_c$ is the coherence
length.

We measure the strength of the Zeeman term via a dimensionless
parameter $Z = \mu B_0 / 2\pi T_c$,\cite{ada05} so that the Zeeman splitting
of the energy levels in dimensionless units is
$\mu B/(2\pi T_c)= Z \: B/B_0$.
Since we expect the orbital critical field $H_{c2}^{orb}$ to be of
order  $B_0$, and the Pauli limit to correspond
to $\mu H_P \sim \Delta_0 \sim T_c$, we find
$H^{orb}_{c2} / H_P \sim Z$.

We first compute the upper critical field by solving the
linearized, with respect to the order parameter, quasiclassical
equations.  Under the combined effect of the orbital and Zeeman
field the transition for strong enough Pauli term becomes first
order at low temperatures, $T<T^*$.
We determine $T^*$ in the
clean limit by evaluating where the $\Delta^4$-term coefficient in
the free energy expansion becomes negative.
The general free energy expansion can be obtained from Eqs.~(\ref{eq:gs}),
but it is complicated in a dirty $d$-wave superconductor\cite{AVorontsov:FFLOimp},
and since the precise location of $T^*$ is
not important for this work we will not look for it here.
At lower temperatures,
the transition line can only be determined from the full free
energy functional, which needs to also account for the possible
existence of the additional modulations in the
Fulde-Ferrell-Larkin-Ovchinnikov (FFLO)
phase~\cite{ada03,ike04,won04a}. While there are many indications
that in CeCoIn$_5$ a new phase exists in the high-field,
low-temperature
range~\cite{ABianchi:2003,cap04,wat04,mar05,VMitrovic:2006,YMatsuda:2007}
the experiments on the vortex state anisotropy are generally
carried out away from that range. Consequently, we do not consider
the FFLO-like modulation here.

\begin{figure}[t]
\centerline{\includegraphics[height=5.5cm]{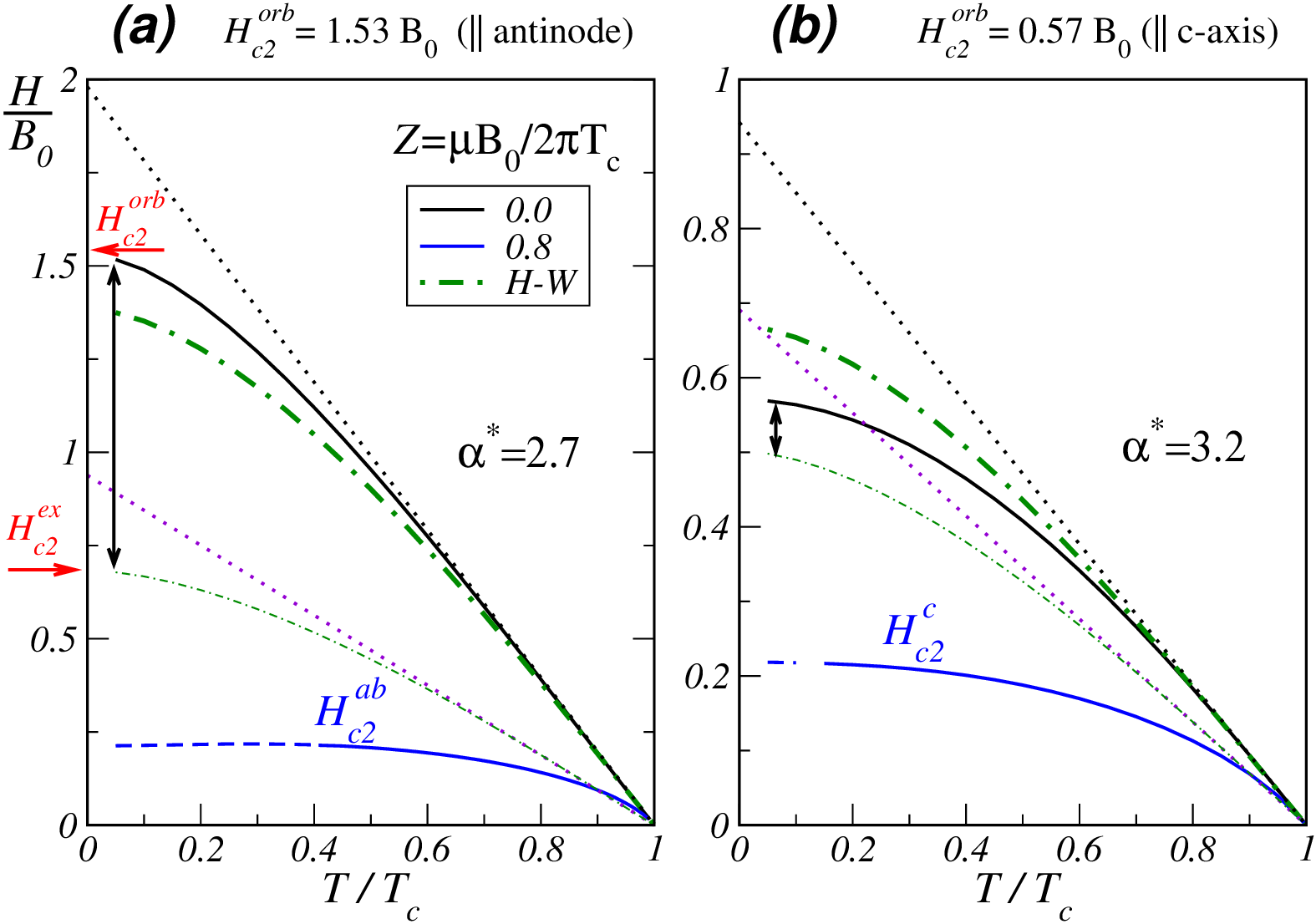}}
\caption{(Color online) Illustration of the possible pitfalls in
the experimental determination of orbital limit on upper critical
field, $H^{orb}_{c2}$, in strongly paramagnetic superconductors.
(a) in-plane field, (b) field normal to the planes. Lower (blue)
solid lines denote $H_{c2}$ for the paramagnetically limiting case
$Z=0.8$, and the dashed lines mark regions of the first-order
transition. ``Turning off'' Zeeman effect ($Z=0$) results in true
orbital $H_{c2}$, the upper (black) solid lines. Dotted lines:
slope $dH_{c2}/dT$ at $T_c$ determined from the points
$T/T_c=(0.9,\;1)$. Thick dot-dashed lines: upper critical field
obtained using the Helfand-Werthamer (HW) result~\cite{hel66} for
the given slope $dH_{c2}/dT$ at $T_c$. Thin dot-dashed lines: HW
profile of $H_{c2}(T)$ for the paramagnetically limited case
$Z=0.8$, resulting in incorrect extracted orbital limit,
$H_{c2}^{ex}$.  Vertical arrows indicate potential discrepancy
between the approximately determined and realistic orbital upper
critical field, see text for details. } \label{fig:hc2ex}
\end{figure}

Fig.~\ref{fig:hc2ex} shows the upper critical field, $H_{c2} (T)$
for a pure system for two different field orientations ($c$-axis
and the in-plane along the gap maximum) and for different strength
of the Zeeman splitting.  Our results for $Z=0$ resemble the
classical results by Helfand and Werthamer \cite{hel64,hel66} (HW) with
differences in profiles attributed to anisotropic gap and
non-spherical FS. The important observation is that the shape of
the temperature dependence of the upper critical field changes
with increasing $Z$, and that the linear region of $H_{c2}(T) \sim
1-T/T_c$ near $T_c$ rapidly shrinks, eventually leading to $H_{c2}
\sim \sqrt{1-T/T_c}$ dependence characteristic of the Pauli
limited field. Hence, as is shown in Fig.~\ref{fig:hc2ex} a
frequently used experimental estimate of the $H^{orb}_{c2} (T=0)$
based on the slope of the measured upper critical field near $T_c$
is not reliable for paramagnetically limited superconductors: that
estimate changes on approaching the Pauli limit.
First, the slopes $dH_{c2}/dT$ of the purely orbital and
paramagnetically limited cases, while equal asymptotically
as $T\rightarrow T_c$, are different when determined within a
reasonable experimental window as shown by dotted lines based on
the values of $H_{c2}$ at $T/T_c=(0.9,\;1)$. We find that even for
moderate paramagnetic coupling a reasonable estimate of the slope
may be obtained only within a window of 1-3\% near $T_c$. Second,
the $H_{c2}(T)$ profile for our chosen Fermi surface and the
$d$-wave order parameter even for the purely orbital case is
different from that found using the HW~\cite{hel66} result for
$s$-wave superconductor with a spherical FS with the same
$dH_{c2}/dT \,(T_c)$. Consequently, for $Z \sim 1$ the estimate of
$H_{c2}^{orb}$ using the HW profile and an approximate slope
$dH_{c2}/dT \,(T_c)$ based on points more than a few percent
away from $T_c$ (thin dot-dashed lines) may significantly
underestimate the magnitude of $H_{c2}^{orb}$, as indicated by the
vertical arrows in Fig.~\ref{fig:hc2ex}.

\begin{figure}[t]
\centerline{\includegraphics[height=5.5cm]{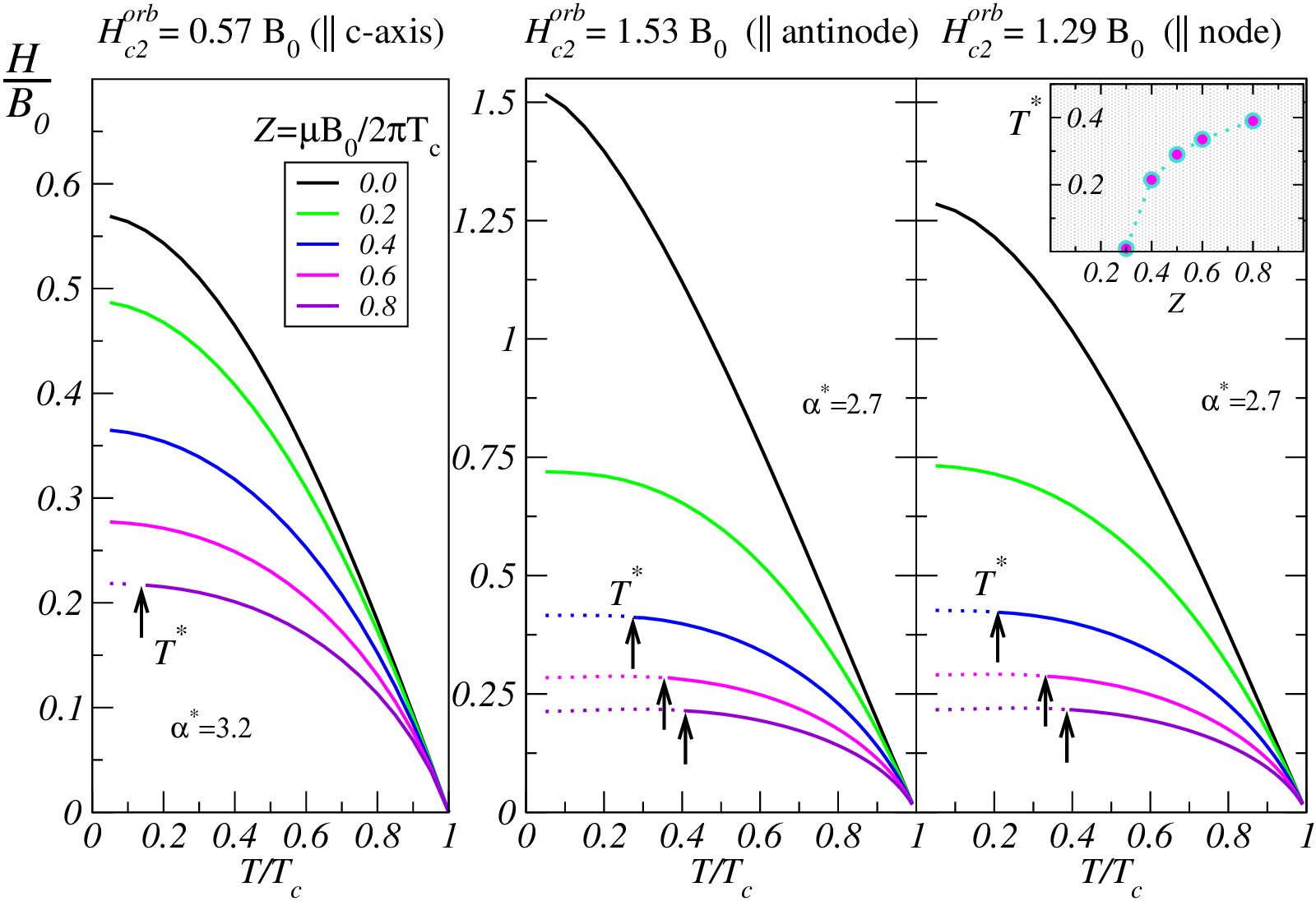}}
\caption{(Color online)
Upper critical field of a pure $d_{x^2-y^2}$ superconductor
with quasi-cylindrical FS for three orientations of the field.
The critical value of the Maki parameter
is about $\alpha^* \sim 3$ for all directions (see text).
The dotted lines indicate first-order transition below $T^*$.
Inset in the right-hand panel: $T^*$ as a function
of the relative strength of the Pauli limiting.
For purely Pauli-limited case $T^*_P \approx 0.56 T_c$. }
\label{fig:hc2}
\end{figure}

The change in the shape of the $H_{c2}(T)$ curve with
increasing paramagnetic contribution is made even more explicit
in Fig.~\ref{fig:hc2}, where we also include the upper critical
field for the in-plane field along the nodal direction. As can be
expected in a magnetically isotropic system, increased Pauli
limiting reduces the anisotropy of the upper critical field caused
by both the in-plane anisotropy of the order parameter (node vs.
anti-node) and the the anisotropy of the Fermi surface ($c$-axis
vs. $a$-$b$ plane).

Finally, in the inset of Fig.~\ref{fig:hc2} we show the onset
temperature for the first order transition, $T^*$ as a
function of the Pauli limiting parameter $Z$. It is instructive to
recast this analysis in terms of the so-called Maki parameter,
$\alpha = \sqrt{2} H^{orb}_{c2}/H_P$. It is conventionally assumed
that the critical value of the Maki parameter that defines a
strong paramagnetic limit is $\alpha > \alpha^* = 1$, as obtained
for the onset of the first order transition in $s$-wave
superconductors with a spherical Fermi surface~\cite{mak66}; this
value is commonly used in the analysis of anisotropic strongly
correlated materials as well. In reality the critical value
$\alpha^*$ depends on details of the Fermi surface and the
superconducting state. For our model $H_{c2}^{orb} \sim 1.4$ in
the $a$-$b$ plane, and $\mu H_P \approx 1.11 \Delta_{d0}/\sqrt{2}$.
Here $\Delta_{d0}= 0.241 (2\pi T_c)$ is the amplitude of the order
parameter at $T=H=0$\cite{vor05b} and the increase of $H_P$ by
factor $1.11$ compared with the similar $s$-wave expression is
caused by the gain in magnetic energy of the $d$-wave
superconducting state by magnetization of the nodal
quasiparticles~\cite{vor06c}. The critical value of the Maki
parameter where first order transition appears is therefore
$\alpha^* = \sqrt{2} H^{orb}_{c2}/H_P \approx 7.0
(H^{orb}_{c2}/B_0) Z^*$,  and, since in-plane $Z^* \sim 0.3$ we have
$\alpha^*_{d-wave} \approx 3.0$, significantly exceeding the value
of unity expected for {\it dirty} isotropic systems.

\begin{figure}[t]
\centerline{\includegraphics[height=5.5cm]{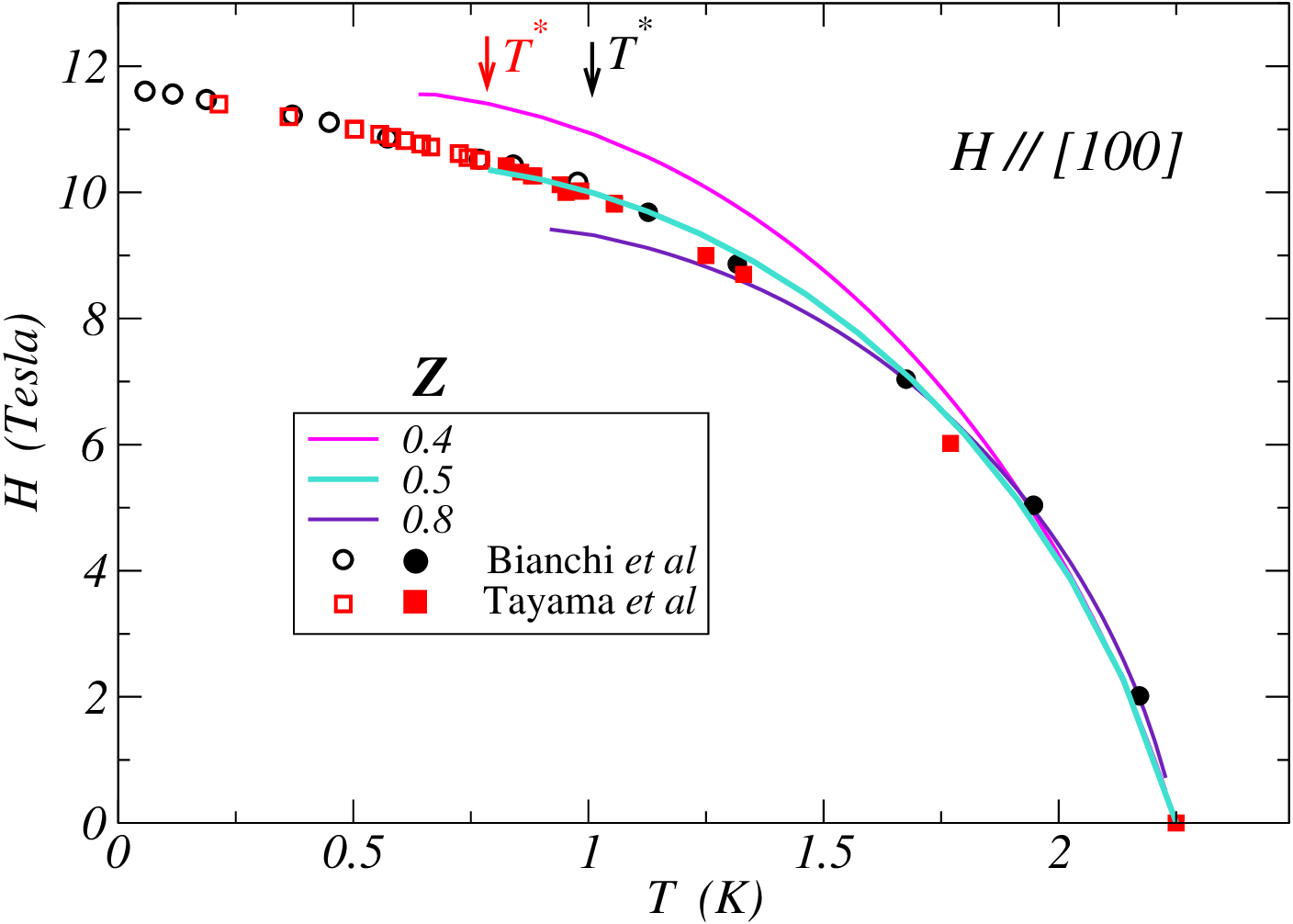}}
\caption{(Color online) A fit of the experimental
$H_{c2} \, // \, [100]$ in CeCoIn$_5$. The data are taken from measurements
of magnetization by T.~Tayama \et\cite{TTayama:2002} and specific
heat by A.~Bianchi \et\cite{ABianchi:2003}. We find that $Z=0.5$
gives the best fit of second order transition line and the onset
of the first order transition $T^*$. } \label{fig:hc2fit}
\end{figure}

We apply our results to CeCoIn$_5$ to analyse the behavior of the
upper critical field along $a$ and $c$ axes. The fit for the
$[100]$ direction is shown in Fig.~\ref{fig:hc2fit}. The best fit
of the second order transition and the location of $T^*$ in the
experimental range is given by $Z^a \sim 0.5$. For $H || c$ the
best fit is given by $Z^c \sim 1.15$. From these two fits we
obtain the value for $B_0$ which is approximately $30 \: Tesla$
for both fields orientations, which is an indication that our
choice of the FS parameters was reasonably good to describe this
system. From this value we find the Fermi velocity $v_f \sim 0.6
\cdot 10^6 \: cm/s$, which is sensible for a heavy fermion system
and agrees with interpretation of the experimental
data~\cite{RMovshovich:2001}, and disagrees with the value $\sim
10^8 \: cm/s$ obtained by authors of Ref.\onlinecite{won04a}. We
then extract the values of the effective electron moment,
$\mu/\mu_B = Z (2\pi k_B T_c)/(\mu_B B_0)$ and obtain
$\mu_{ab}/\mu_B \approx 0.35$ and $\mu_{c}/\mu_B \approx 0.7$, in
agreement with Ref.\onlinecite{won04a} where the varying
parameters were Fermi velocity and the $g$-factor.

We do note that in the current model the orbital
pairbreaking is stronger for the nodal direction,
Fig.~\ref{fig:PD}(a), hence the Pauli effect is relatively less
important, and the range of first order transition is smaller for
that orientation, (Fig.\ref{fig:PD}(c)). However, this particular
aspect depends sensitively on the in-plane shape of the Fermi
surface. We took the Fermi surface to be rotationally symmetric,
and the question of how this conclusion is affected by a more
realistic Fermi surface shape is left for future studies.

\section{Transport and specific heat anisotropy}

We are now in the position to analyse how the Pauli limiting
affects the measured anisotropies of the specific heat,
$C(\phi_0)$, and longitudinal thermal conductivity,
$\kappa_{xx}(\phi_0)$, when magnetic field is rotated in the basal
plane with respect to the crystalline axes. The direction of the
field, $\phi_0$, is measured from the gap's maximum along the
$a$-axis. As is well known, both quantities show oscillations,
with either minima or maxima when the field is aligned with the
nodal
directions~\cite{IVekhter:1999R,AVorontsov:2006,ABVorontsov:2007dos,ABVorontsov:2007kap,YMatsuda:2006,TSakakibara:2007}.
The location of the inversion line in the $T$-$H$ plane that
separates the regions with minima and maxima indicating the gap
nodes is of exceptional experimental relevance as it affects the
conclusions about the shape of the gap in a given compound.

\begin{figure}[t]
\centerline{\includegraphics[height=5.5cm]{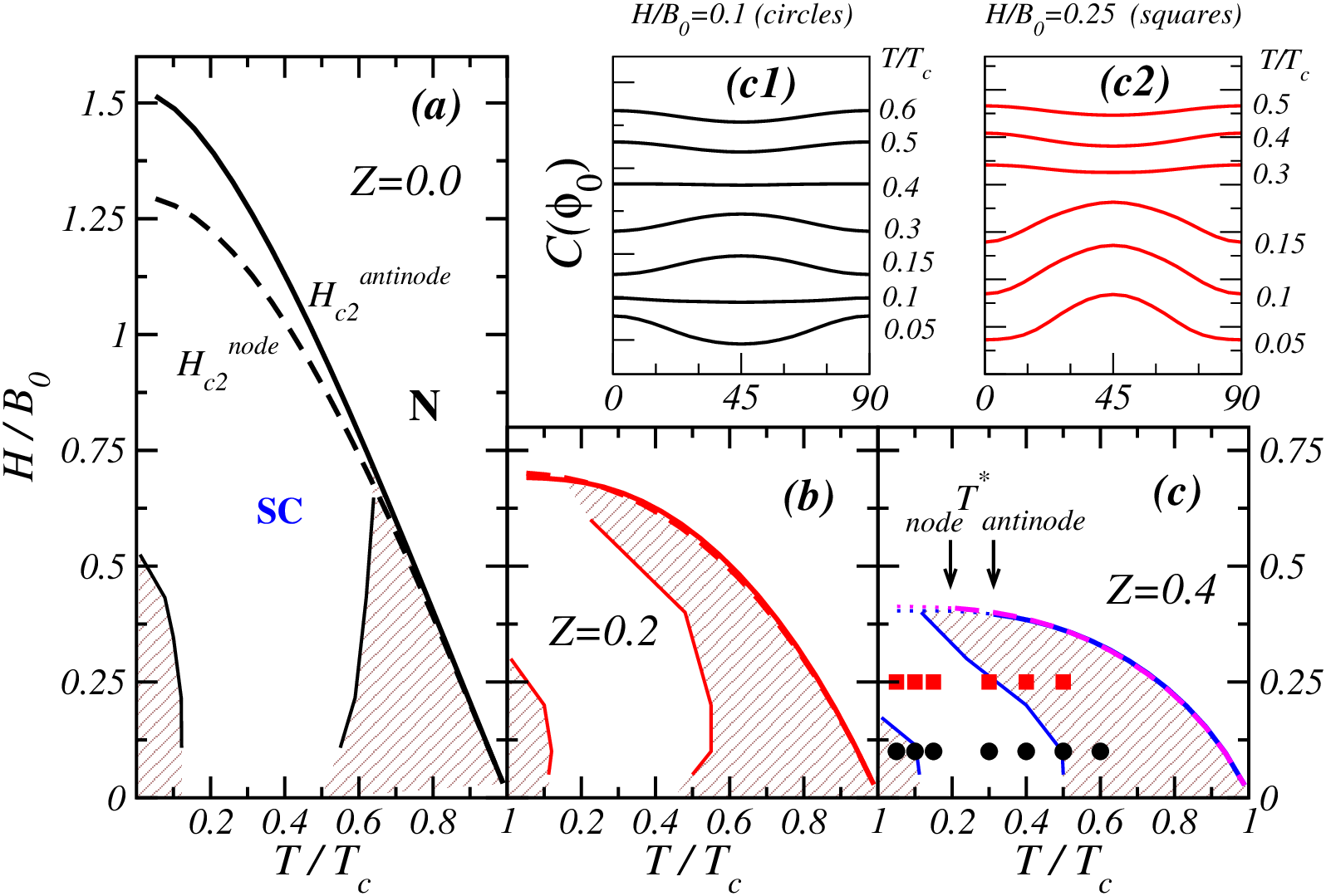}}
\caption{(Color online) The anisotropy of heat capacity phase
diagram for three Zeeman couplings $Z=0.0$ (a), $Z=0.2$ (b) and
$Z=0.4$ (c). Shaded areas correspond to minimum of the heat
capacity for $\vH || (node)$. Panels (c1) and (c2) demonstrate how
anisotropy of heat capacity, $C(\phi_0)$, varies with temperature
and magnetic field for $Z=0.4$. The curves for different
$T/T_c$ are shifted vertically for clarity. } \label{fig:PD}
\end{figure}

The results are most clearly presented in the form of a phase
diagram in the $T$-$H$ variables that indicates each of the
regions. In Fig.~\ref{fig:PD} we show the evolution of such phase
diagram for the specific heat anisotropy with increasing Zeeman
coupling. We compute the density of states, $N(\vare,
T)=-(N_0/2\pi)\langle\Im g(\hvp,\vare)\rangle_{\hvp}$, where
inclusion of $T$ signifies that the self-consistently determined
gap is temperature-dependent, obtain the entropy,
\begin{eqnarray}
S(T,\bm H)&=&-\int_{-\infty}^{\infty} d\omega N(\vare, T)
\biggl[f(\vare)\ln f(\vare)\nonumber\\&& \qquad\qquad +
(1-f(\vare))\ln(1-f(\vare))\biggr]\,, \label{entropy}
\end{eqnarray}
and numerically differentiate it to find $C/T$. The result for
purely orbital coupling to magnetic field, $Z=0.0$ in
Fig.\ref{fig:PD}(a) is in agreement with
Ref.~\onlinecite{ABVorontsov:2007dos}: In the two shaded regions
(near $H_{c2}$ and in low-$T$ low-$H$ corner) the heat capacity
attains its minimum when the field is along the nodal directions
of the gap. Clearly, since the anisotropy in the heat capacity is
field-induced,  the amplitude of the anisotropy is very small at
low fields for all temperatures, and extremely small near $T_c$.
Hence the challenge is to go to sufficiently high fields to see
the signal while knowing whether the maximum or a minimum of the
anisotropic signal corresponds to the nodal directions.

As the relative strength of the spin coupling increases,
Fig.~\ref{fig:PD}(b) and (c), we note that the ``nodal minimum''
region near $H_{c2}$ grows at the expense of the intermediate
field unshaded region. This is concomitant with the disappearance
of the anisotropy  of the upper critical fields for nodal and
antinodal directions. An obvious conjecture is that for $Z=0$ at
intermediate temperatures and {\em moderately high fields} it is
the anisotropy of the upper critical field that controls the
anisotropy of the specific heat: since
$H_{c2}^{node}<H_{c2}^{antinode}$, at a fixed external field
$H/H_{c2}^{node}>H/H_{c2}^{antinode}$, and the density of states,
along with the heat capacity, is higher for the field along the
nodal direction. As the Zeeman coupling is increased and the
critical fields along the nodal and the antinodal directions
become nearly equal, this unshaded region shrinks. In
Fig.~\ref{fig:PD}(c) we tuned $Z>Z^*$, and see that the
critical fields for the two in-plane directions are almost
indistinguishable, and that there is a region of first order
transition at low temperatures $T<T^*$, indicated by the arrows.

Importantly for our purposes, the low-$T$, low-$H$ shaded region
(minima for the field along the nodes) remains mostly unchanged
with increasing $Z$, terminating at $T/T_c \sim 0.1$ and
$H/H_{c2}^Z \sim 0.4$. This region is still dominated by the nodal
quasiparticles and ``semiclassical'' physics, $N(0,H) \sim
\sqrt{H/H_{c2}^{orb}}$,~\cite{GVolovik:1993,MineevSamokhin}
compared to a linear contributions due to Zeeman shifts (for these
$Z$'s) and vortex cores.

The insets Figs. \ref{fig:PD}(c1) and \ref{fig:PD}(c2) provide
image of the heat capacity anisotropy for $Z=0.4$ coupling for two
different fields at locations indicated by circles and squares
correspondingly. The heat capacity expression that we use to plot
the angular dependence,
    \be
    C(T,\vH) = \int_{-\infty}^\infty {d\vare \, \vare^2 \over 4
T^2} {N_\uparrow(\vare,T,\vH) + N_\downarrow(\vare,T,\vH) \over
\cosh^2(\vare/2T)}\,,
    \ee
is, strictly speaking, valid only at low temperature when the
order parameter is essentially temperature-independent.  The
difference between this expression and the exact result obtained
from the entropy differentiation is small far from the
superconducting transition, see
Refs.~\onlinecite{AVorontsov:2006,ABVorontsov:2007dos}.

\begin{figure}[t]
\centerline{\includegraphics[height=4cm]{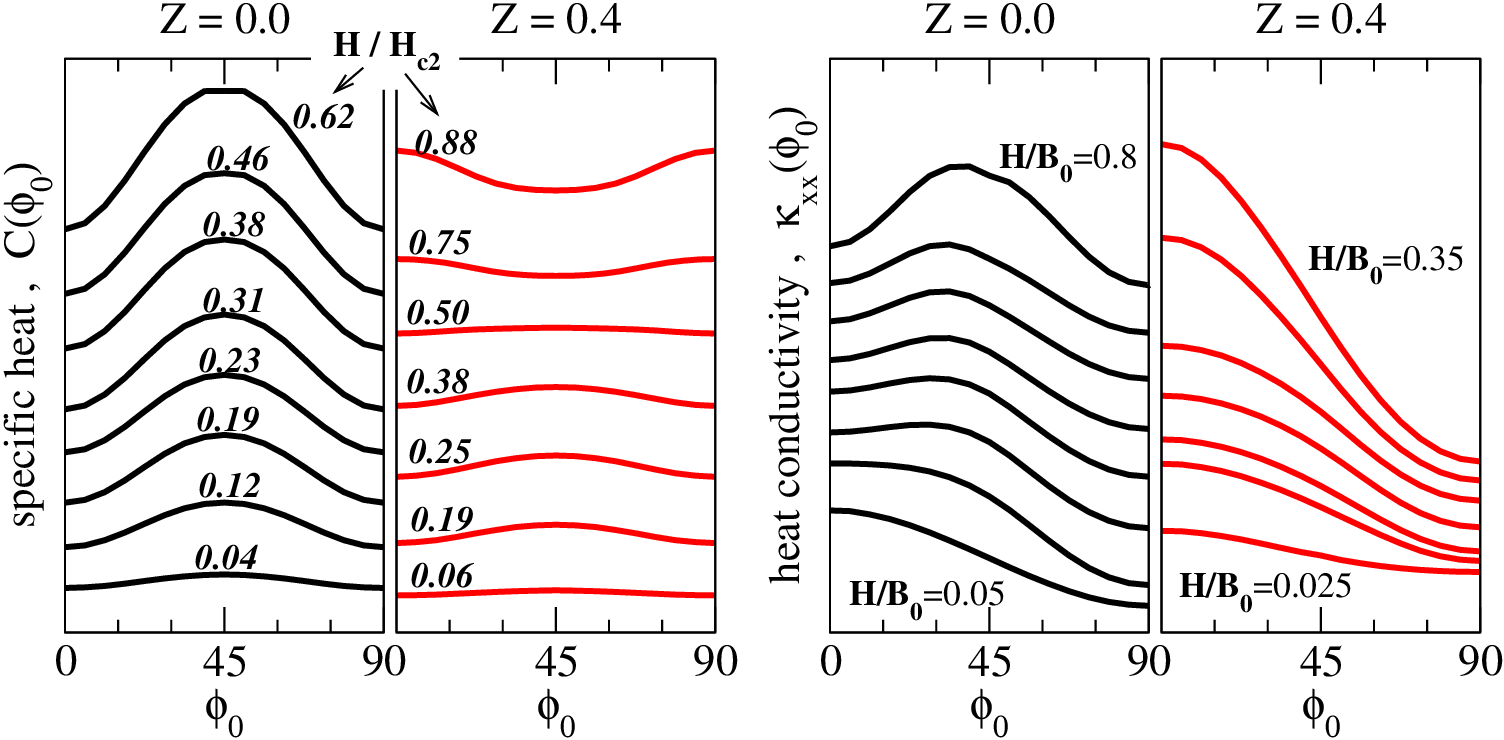}}
\caption{(Color online) Anisotropy of the specific heat (left) and
heat conductivity (right) for $Z=0.0$ and $Z=0.4$ at $T/T_c=0.3$
and various fields. The field values are shown on the left in
terms of $Z$-dependent $H_{c2}$. The curves were shifted
vertically for clarity. The absolute values of $C$ and $\kappa$
and their anisotropies presented in Figures below. The range of
fields in $H/B_0$ for convenience indicated on the right. }
\label{fig:anis}
\end{figure}

We also compute the anisotropy of the thermal conductivity along
the crystalline $x$ direction for unitarity scattering (phase
shift $\pi/2$) and the normal state scattering rate
$\Gamma/2\pi T_c = 0.007$. The thermal conductivity is calculated on equal
footing with the density of states and
is given by~\cite{IVekhter:1999,ABVorontsov:2007kap}
\begin{eqnarray}
\frac{\kappa_{xx}(T,H)}{T} &=& \int\limits^{+\infty}_{-\infty} \;
\frac{d\vare}{2 T} \frac{\vare^2}{T^2} \cosh^{-2}\frac{\vare}{2T}
\\
\nonumber &\times& \Big< v_{f,x}^2  \, N(T, \bm H; \hat{\vp},
\vare) \; \tau_H(T, H; \hat{\vp},\vare)\Big>_{\hvp} \,,
\end{eqnarray}
with the effective scattering rate
    \be
    \frac{1}{2\tau_H} =
    - \Im \Sigma^R + \sqrt{\pi}{2 \Lambda \over |\tilde{v}_f^\perp|}
    \frac{\Im[g^R \, W(2\tilde{\vare}\Lambda/|\tilde{v}_f^\perp|)]}
    {\Im \, g^R} |\Delta_0 \cY|^2
    \,.
    \label{eq:tauH}
    \ee
Here the label $R$ denotes a retarded function, and we use the
angle-resolved density of states.

Fig.~\ref{fig:anis} shows characteristic profiles of the heat
capacity and heat conductivity when the field is rotated in the
basal plane. Note that in the left panel we labeled the fields
according to the ratio $H/H_{c2}$: the upper critical field
changes between $Z=0$ and $Z=0.4$, so that the effective field
range is different. As discussed above, the major difference
between two values of the Zeeman splitting is that for higher $Z$
there exists a region of shallow minimum at the node ($45^\circ$ in our case)
for high fields.

The second panel of Fig.~\ref{fig:anis} shows the behavior of the
thermal conductivity for the same fields, now labeled in units of
$B_0$. Qualitatively, the peak in the angle-dependent $\kappa$ for
$Z=0$ at $45^\circ$ disappears at $Z=0.4$, making the dependence
of the thermal conductivity more twofold.

\begin{figure}[t]
\centerline{\includegraphics[width=0.9\columnwidth]{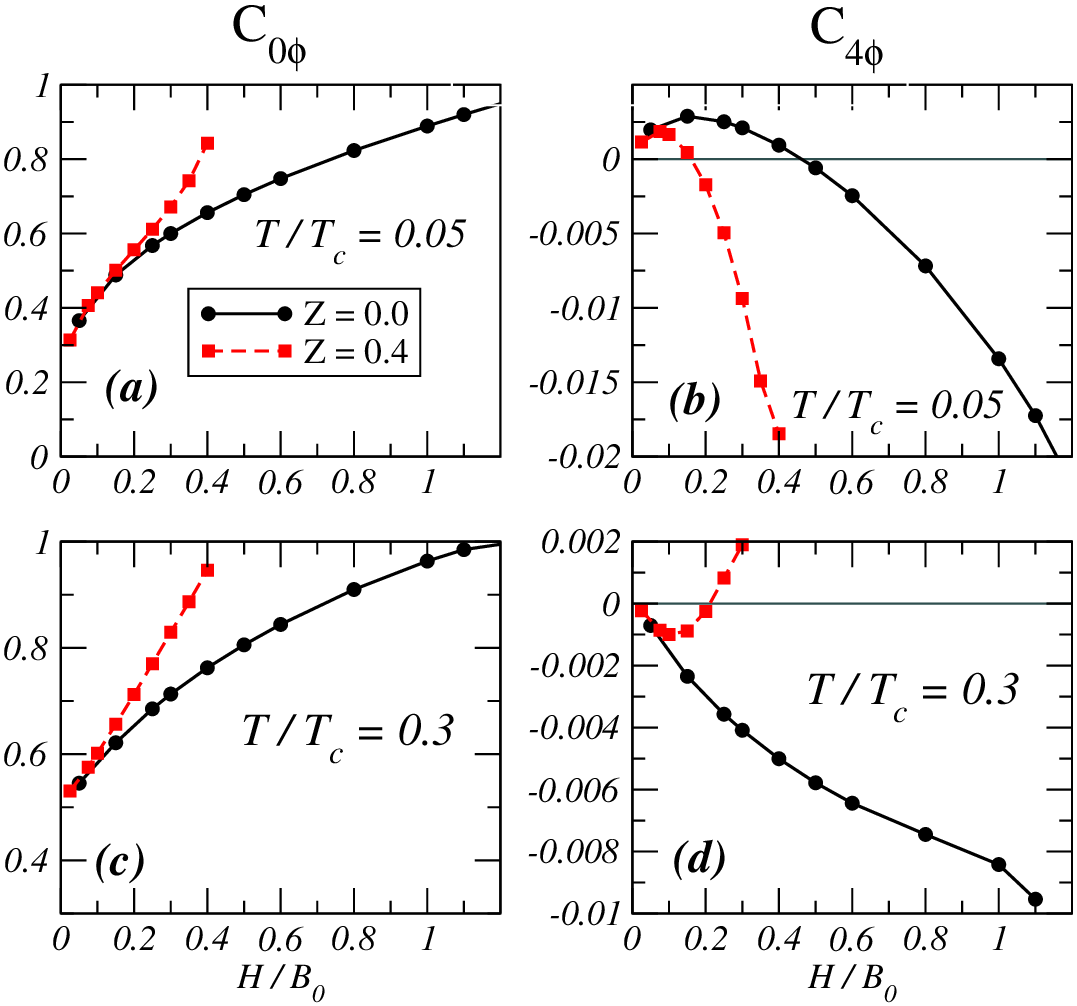}}
\caption{(Color online) $C_{0\phi}$ and $C_{4\phi}$ coefficients
in heat capacity expansion at two different temperatures,
$T=0.05T_c$ and $T=0.3T_c$. Positive four-fold coefficient means
that the specific heat has minimum for the field along a node in
the gap, while negative $C_{4\phi}$ corresponds to the maximum for
the field along a node. } \label{fig:coefC}
\end{figure}
To quantify these trends we follow the approaches taken in
experiment and expand both quantities in harmonics of the angle
between the field and the heat current,
    \bea
    \frac{C(\phi_0,T,H)/T}{(C/T)_N} &=
&C_{0\phi} + C_{4\phi} \cos4\phi_0 \,, \label{eq:coefC}
\\
\nonumber \frac{\kappa_{xx}(\phi_0,T,H)/T}{(\kappa_{xx}/T)_N} &=&
\kappa_{0\phi}  + \kappa_{2\phi} \cos2\phi_0 + \kappa_{4\phi}
\cos4\phi_0 \,, \label{eq:coefK}
    \eea
and present the evolution of different coefficients with field and
temperature.

\begin{figure}[t]
\centerline{\includegraphics[width=0.9\columnwidth]{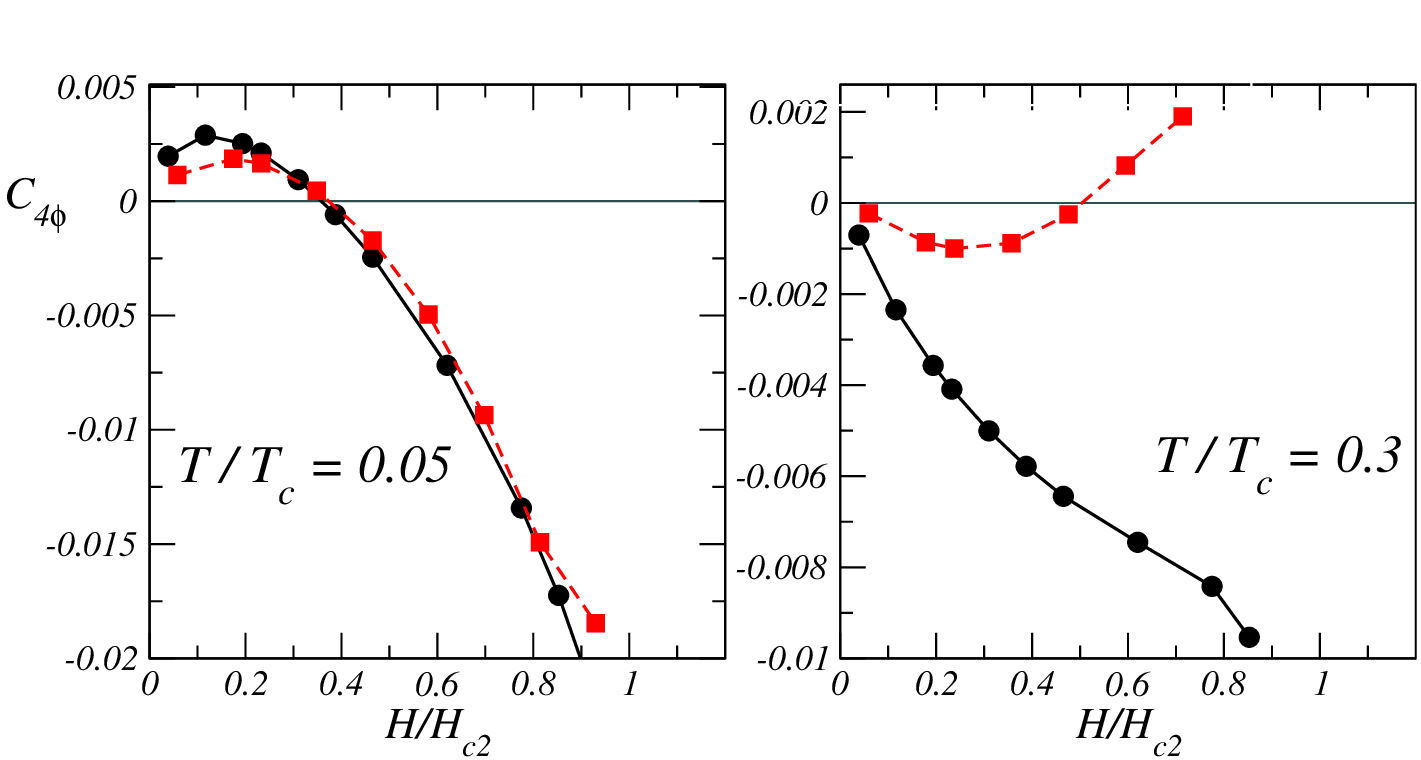}}
\caption{(Color online) The anisotropic component of the specific
heat at temperatures $T=0.05T_c$ (left panel) and $T=0.3T_c$
(right panel) for $Z=0$ and $Z=0.4$.  } \label{fig:C4vsHc2}
\end{figure}

First, focus on the behavior of the specific heat as shown in
Fig.~\ref{fig:coefC}. At low fields for $Z=0$ the isotropic part,
$C_{0\phi}$,  shown in panels (a) and (c), exhibits the
approximate $\sqrt{H}$ behavior expected of a nodal
superconductor~\cite{GVolovik:1993,CKubert:1998SSC}. With
increased Zeeman contribution this component of the specific heat
becomes more linear in field~\cite{yan98,IVekhter:2001}. This is,
of course, simply because the spin-split quasiparticle spectra
produce a density of states at zero energy, $N(0,H) \simeq (\mu
H)/\Delta_0$, with the prefactor that, for high $\mu$, can be
large enough to dominate the sublinear Volovik term already at low
fields. At low temperatures $C_{0\phi}$ acquires a positive
curvature at high fields, in agreement with
Ref.~\onlinecite{ada05}.

The qualitative behavior of the anisotropic term,, $C_{4\phi}$, is
similar for the two values of $Z$ at low temperature, panel (b),
but is distinctly smaller at higher temperature, shown in panel
(d), and even changes sign near $H_{c2}$. The comparison becomes
even more clear if the amplitude $C_{4\phi}$ is plotted against
the upper critical field for given values of Zeeman splitting, as
in Fig.~\ref{fig:C4vsHc2}. At low temperature there is essentially
no difference between the purely orbital case and that of a
moderately strong Zeeman coupling, and hence in this regime the
anisotropy depends only on the shape of the gap and the underlying
Fermi surface. In contrast, at $T/T_c=0.3$ for the case of weak
Zeeman splitting the coefficient does not change sign, as seen
already from Fig.~\ref{fig:PD}(a). In contrast, for $Z=0.4$ as
shown in Fig.~\ref{fig:PD}(c), there is a change of sign in the
anisotropy of the specific heat, due to the ``isotropization'' of
the upper critical field, as discussed above.

\begin{figure}[t]
\centerline{\includegraphics[width=0.9\columnwidth]{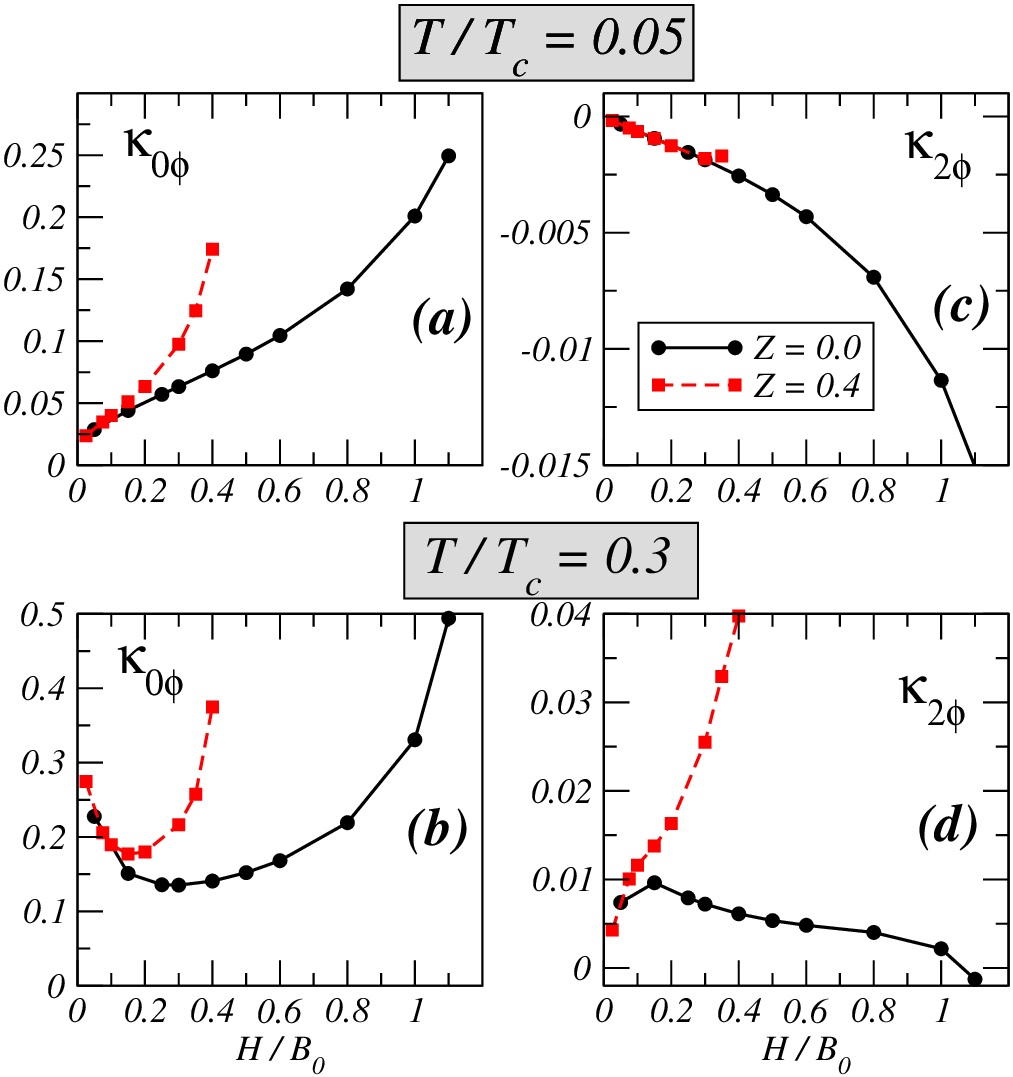}}
\caption{(Color online) Field dependence of the coefficients
$\kappa_0$ and $\kappa_2$ in the expansion of the thermal
conductivity, Eq.(\ref{eq:coefK}) at $T=0.05T_c$ (top two panels)
and $T=0.3T_c$ (bottom panels) for $Z=0$ and $Z=0.4$.}
\label{fig:kappa02}
\end{figure}
\begin{figure}[b]
\centerline{\includegraphics[width=0.9\columnwidth]{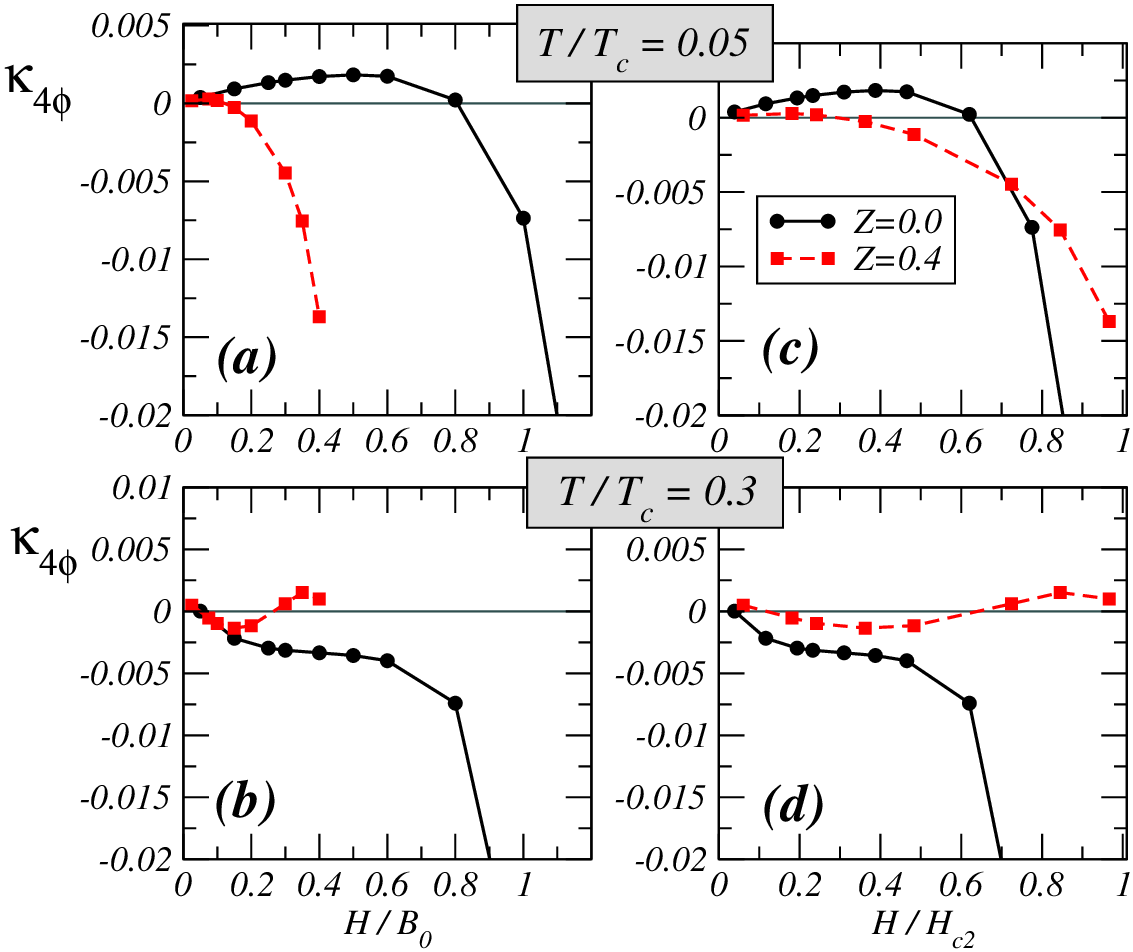}}
\caption{(Color online) The anisotropic fourfold term of the
thermal conductivity in Eq.(\ref{eq:coefK}) is shown for
$T=0.05T_c$ in panels (a) and (c), and for $T=0.3T_c$ in panels
(b) and (d) as function of the field, panels (a) and (b) and the
reduced field, $H/H_{c2}$, panels (c) and (d).} \label{fig:kappa4}
\end{figure}

The same trend is seen in the thermal transport anisotropy. The
behavior of the average thermal conductivity, $\kappa_0$ and the
twofold term responsible for the difference between heat transport
parallel and perpendicular to the vortices, is shown in
Fig.~\ref{fig:kappa02}. At low fields the scattering of the
quasiparticles on the vortices is determined by the vortex
concentration, $n\sim H/\Phi_0$, and therefore, at low
temperature,  only weakly depends on the Zeeman field. As the
temperature increases, however, the  differences between the two
cases become much more pronounced. In particular, it is worth
noting that the two-fold symmetry $\kappa_{2\phi}>0$
($\kappa_{xx}(\vj_h || \vH) > \kappa_{xx}(\vj_h \perp \vH)$) is
much more pronounced in the Pauli-limited case.

The fourfold ``nodal'' term, Fig.~\ref{fig:kappa4}, shows the
behavior broadly similar to that of the anisotropic component of
the specific heat. The low-temperature behavior as a function of
the reduced field, $H/H_{c2}$ is similar for the cases of weak and
strong Pauli limiting, while the behavior at moderate temperatures
is very different starting at moderate fields. Once again, the
coefficients for $Z=0$ and $Z=0.4$ have the opposite signs
starting at  $H\sim 0.5 H_{c2}$.

\section{Conclusions}

\begin{figure}[t]
\centerline{\includegraphics[width=0.9\columnwidth]{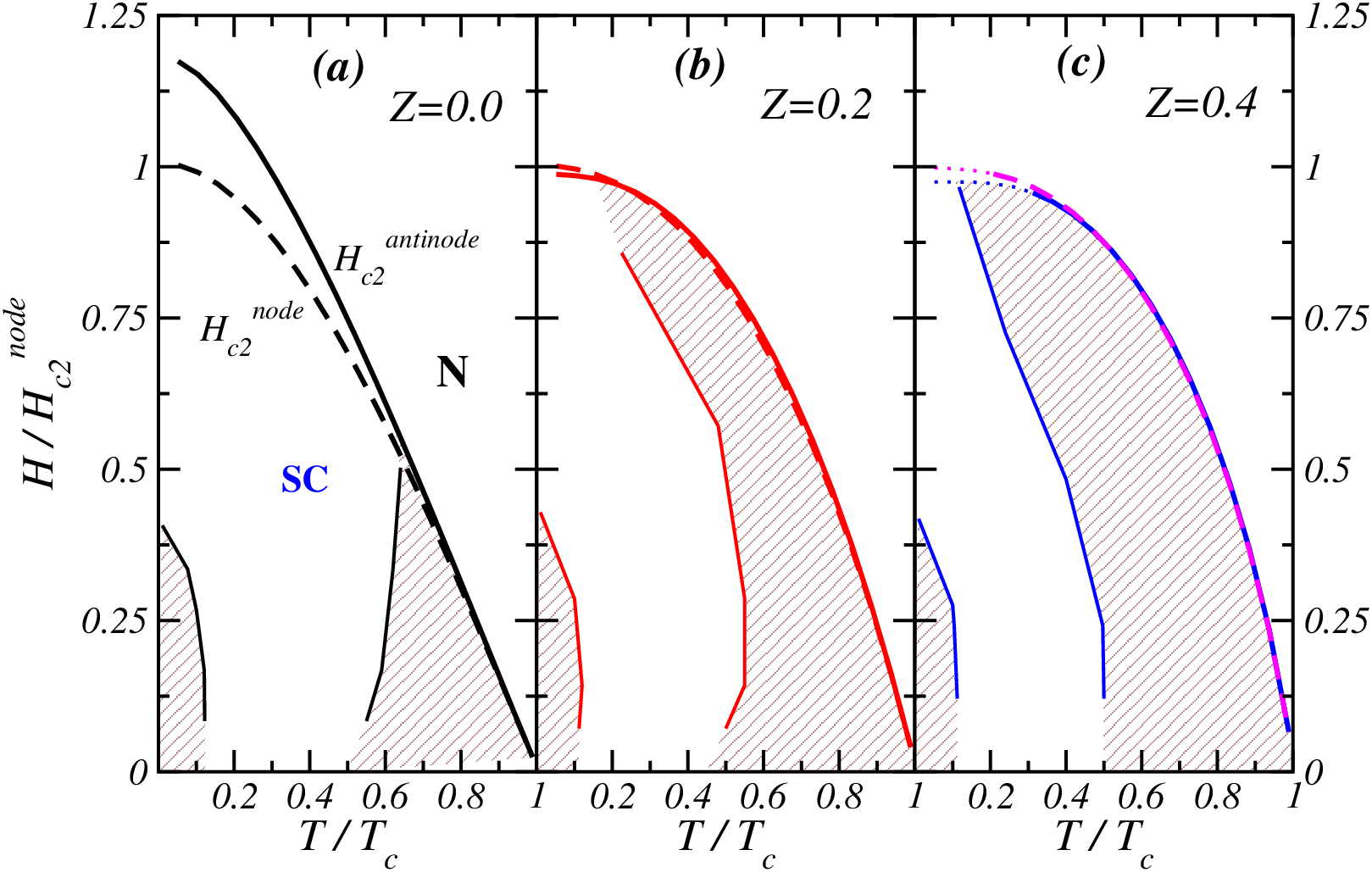}}
\caption{(Color online) The phase diagram  for the anisotropy of
the specific heat under rotated magnetic field for different
strength of Pauli pairbreaking term. Shaded areas correspond to
minimum of the heat capacity for $\vH || (node)$. Notice that the
shaded region at low $T$ and $H$ is identical in all three
panels.} \label{fig:PDsummary}
\end{figure}
In conclusion, we calculated in a $d$-wave superconductor with
quasi-cylindrical FS the $H_{c2}$ and its evolution for transition
between orbital and Pauli limits. We find that in this system the
critical value of the Maki parameter for paramagnetic limit
criterion is $\alpha^*=3$. That values depends on both the
symmetry of the order parameter and the shape of the Fermi
surface, and hence there is no universal threshold value for that
parameter that ensures the first order transition in a given
system. We also find that the linear extrapolation of the upper
critical field in Pauli-limited superconductors from the vicinity
of $T_c$ to low temperatures does not accurately predict the value
of the orbital critical field.

We showed that moderately large Zeeman contribution does not alter
the typical behavior of $C-$ and $\kappa-$anisotropy at low
temperatures and fields. This is summarized in the phase diagram
in Fig.~\ref{fig:PDsummary}. The region at low $T$ and $H$ where
the minima in the fourfold terms in the heat capacity and thermal
conductivity occur for the field along the nodes is essentially
insensitive to the strength of the Zeeman coupling. The
differences, of course, occur at higher temperatures and fields.

This finding proves that, even in systems with strong paramagnetic
contribution where $H_{c2}$ anisotropy in the $a$-$b$ plane is
largely absent and therefore cannot be used to infer the nodal
structure~\cite{wei06}, the anisotropy of heat conductivity and
heat capacity at moderate to low $H$ and $T$ still can help
determine the location of the nodes of the order parameter on the
Fermi surface. The previous analysis~\cite{AVorontsov:2006} of the
inversion of the anisotropy in CeCoIn$_5$ remains valid when the
Zeeman term is accounted for, and unequivocally indicates the
$d_{x^2-y^2}$ shape of the order parameter. This also provides
further support for the interpretation of
Ref.~\onlinecite{KAhn:2009} that confirmed the inversion of the
specific heat oscillations.

\section*{Acknowledgements}

This research was supported in part by the US Department of Energy
via Grant No. DE-FG02-08ER46492. We are also grateful for the
hospitality of the Aspen Center for Physics, where part of this
work was done.


\end{document}